\documentclass[11pt]{article}

\usepackage[T1]{fontenc}
\usepackage[letterpaper,left=1.02in,right=1.02in,top=1.05in,bottom=1.02in]{geometry}
\usepackage{newtxtext}
\usepackage{newtxmath}
\usepackage{amsmath,bm,mathtools}
\usepackage{array,booktabs,tabularx}
\usepackage{graphicx}
\usepackage{microtype}
\usepackage{siunitx}
\usepackage{xcolor}
\usepackage[font=small,labelfont=bf,labelsep=period,justification=justified]{caption}
\usepackage[backend=biber,style=numeric-comp,sorting=none,maxbibnames=8]{biblatex}
\usepackage{enumitem}
\usepackage{float}
\usepackage[pdfusetitle]{hyperref}
\usepackage{orcidlink}
\usepackage{cleveref}

\hypersetup{
  colorlinks=true,
  allcolors={blue!55!black},
  breaklinks=true
}

\setlist{nosep}

\newcommand{\vect}[1]{\bm{#1}}
\newcommand{\cH}{\mathcal{H}}

\newcommand{\ket}[1]{|#1\rangle}

\DeclareMathOperator{\rank}{rank}

\title{Conservation-Syndrome Quantum Error Correction for Lattice-Boltzmann Quantum Algorithms}
\author{\texorpdfstring{Muhammad Idrees Khan\,\orcidlink{0009-0003-0046-821X}\\
Independent researcher}{Muhammad Idrees Khan}}
\date{}

\begin{document}
\maketitle

\begin{abstract}
Reliable multistep quantum lattice-Boltzmann evolution requires controlling computational faults. When ideal collision preserves encoded mass and momentum exactly, changes in those charges can provide syndromes for selected faults at the post-collision checkpoint. A fault that shifts one population then leaves a unique mass-momentum residual labeled by the lattice velocity. A coherent reference register stores the expected charges, so the same test remains valid while the physical fluid charges vary across the lattice. Single bit flips on population number registers occupy disjoint syndrome sectors and satisfy the Knill-Laflamme condition. A state-vector demonstration on eighteen data qubits of the two-dimensional nine-velocity lattice (D2Q9) recovers that single-bit family to double-precision roundoff. Conservation still leaves a fifteen-dimensional kinetic nullspace on the three-dimensional nineteen-velocity lattice (D3Q19). Exact multiple-relaxation-time (MRT) streaming analysis ranks those charge-preserving modes by the order at which they return to density or momentum. A classical D3Q19 decaying-flow simulation at moderate Reynolds number then injects about $1.75\times10^4$ selected single-population shifts per realization. Recovery of that alphabet returns the trajectory to floating-point roundoff. The quantum statements assume an ideal charge-preserving collision and a reliable reference. The result is an inner recovery map for a stated charge-changing alphabet together with a kinetic classification of the unresolved sector.
\end{abstract}

\section{Introduction}

Lattice-Boltzmann methods (LBM) evolve discrete populations through local collision and lattice streaming \cite{lallemand2000theory,dhumieres2002mrt,succi2018lbm,kruger2017lbm}. Their macroscopic variables are low-order moments of those populations. Collision preserves mass and momentum. Higher kinetic moments relax toward equilibrium. Streaming transports the post-collision populations to neighboring sites. This separation gives LBM a compact kinetic structure that is attractive for quantum algorithms.

At an exact, unforced collision checkpoint, mass and momentum remain fixed even as kinetic moments relax. Some population-register faults violate these equalities, while others leave them unchanged. This raises the question of which faults a local charge residual can identify and coherently correct when the physical fluid charges vary across sites and timesteps.

Quantum lattice-Boltzmann (QLBM) research has focused mainly on collision, streaming, state encoding, and circuit constructions \cite{schalkers2024encoding,duong2026denoising,georgescu2025qlbm,sanavio2024carleman}. Carleman linearization and related methods have then been used for multistep evolution and for dissipative or nonlinear collisions \cite{sanavio2024routes,wang2025nonlinear,bastida2026carleman,wang2026spectralviscosity}. Open-system collision operators and digital number-register timesteps supply further dissipative and nonlinear constructions \cite{khan2026deterministic,khan2026lindblad,khan2026digital}. A learned surrogate collision circuit has also imposed mass conservation and lattice equivariance directly on a two-dimensional nine-velocity (D2Q9) quantum approximation \cite{lacatus2026surrogate}. An end-to-end incompressible LBM algorithm has recently supplied fault-tolerant complexity and gate-count analysis \cite{jennings2026endtoend}. Hardware-oriented work has also addressed noise and resilience in quantum fluid and differential-equation solvers through mitigation and model-specific compensation strategies \cite{tiwari2025realizable,gan2025nonlinear}. Quantum lattice-gas studies have examined noise sensitivity and conserved quantum invariants \cite{bastida2025qlga,fonio2025invariants}.

Quantum error correction (QEC) addresses a different layer. A prescribed error family is correctable when its action on the logical subspace satisfies the Knill-Laflamme relations \cite{knill1997qec}. Physical symmetry can supply useful structure. Symmetry verification rejects components outside a target sector \cite{bonet2018symmetry}. Symmetry operators can define active codes \cite{niu2018symmetry,qin2024symmetry,bradshaw2026symmetry}. Gauge-symmetry codes restore a physical Gauss-law sector in lattice gauge theory \cite{rajput2023gauge,spagnoli2026gauge}. These works establish the broad idea that physics and fault tolerance can share mathematical structure.

We address this question by using those charge residuals as error syndromes. The construction then treats the large family of charge-preserving errors through a second quantity called hydrodynamic visibility. This quantity is the first wave-number order at which a kinetic perturbation reaches density or momentum through streaming.

We develop four linked results. First, the three-dimensional nineteen-velocity (D3Q19) conservation matrix gives an injective syndrome for a selected single-population shift alphabet. Second, a reference-charge quantum encoding turns the relative conservation residual into an active quantum syndrome while allowing the physical mass and momentum to vary across sites. The same construction corrects single population-register bit flips. For a $B$-bit D3Q19 number encoding it yields $19B$ distinct bit-error labels. Third, exact Fourier expansion of the D3Q19 multiple-relaxation-time (MRT) operator used here resolves the fifteen-dimensional conservation nullspace into $O(k)$, $O(k^2)$, and $O(k^3)$ visibility classes. Fourth, a moderate-Reynolds-number D3Q19 MRT decaying isotropic simulation verifies repeated recovery of the selected additive single-shift alphabet in classical floating-point arithmetic.

The quantum recovery statements assume an ideal charge-preserving collision and a reliable reference register. Finite-width population reconstruction must preserve the encoded charges exactly at the syndrome checkpoint.

The located literature contains broad symmetry-based QEC, QLBM noise mitigation, quantum-invariant analyses, and classical LBM resilience. Those works leave open an inner recovery map at the local collision checkpoint for a stated charge-changing family, together with a hydrodynamic visibility classification of the conservation-nullspace population errors. The recovery claim remains tied to the stated error alphabet. Phase faults, leakage, correlated faults, and faults in the syndrome circuitry require an outer protection layer.

\section{Relation to prior fault-tolerance ideas}

Algorithm-based fault tolerance (ABFT) predates quantum computing. Classical ABFT augments numerical data with checksums that propagate through the target algorithm \cite{huang1984abft}. LBM has served as a test problem for grid-sampling dual modular redundancy and checkpoint recovery \cite{ren2015gsdmr}. Sparse-grid combination methods provide another resilience route and have been demonstrated with an LBM application \cite{ali2016sparse}.

Quantum symmetry methods form the closest conceptual prior art. Bosonic symmetry codes use physical symmetry operators in the code construction \cite{niu2018symmetry}. Symmetry verification uses measured conserved quantities for error mitigation \cite{bonet2018symmetry}. Redundant string symmetries support active recovery \cite{qin2024symmetry}. Representation-theoretic constructions extend symmetry-aware codes beyond Pauli groups \cite{bradshaw2026symmetry}. Rajput et al. combine the Gauss law constraint with active bit and phase-flip correction for lattice gauge theory \cite{rajput2023gauge}. Their recovery targets a physical gauge sector. Gauge-covariant codes later embed that symmetry in a fault-tolerant lattice-gauge algorithm \cite{spagnoli2026gauge}. Conservation language also appears in matching-decoder theory \cite{brown2023conservation}. Partial QEC can shape residual logical noise into a desired dissipative channel \cite{dambal2026partial}.

QLBM and quantum lattice-gas work has addressed hardware noise and physical constraints from different directions. Tiwari et al. combine hardware-oriented circuit reduction with noise estimation and mitigation in an experimental QLBM workflow \cite{tiwari2025realizable}. Bastida Zamora et al. quantify the resilience of a quantum lattice-gas algorithm under device noise \cite{bastida2025qlga}. Fonio et al. analyze conserved quantities and quantum invariants in fluid-dynamic lattice-gas automata \cite{fonio2025invariants}. Schalkers and M\"oller encode collision as a unitary mixing of velocity patterns that share the same mass and momentum \cite{schalkers2024encoding}. Those invariants therefore define legal codewords. L\u{a}c\u{a}tu\c{s} and M\"oller constrain a learned quantum Bhatnagar-Gross-Krook (BGK) collision approximation by mass conservation and lattice equivariance while treating momentum conservation through the training objective \cite{lacatus2026surrogate}. Jennings et al. develop an end-to-end incompressible LBM quantum algorithm and analyze its fault-tolerant computational cost \cite{jennings2026endtoend}. Wang et al. use an H-step to control artificial within-node correlations in a nonlinear QLBM construction and identify error correction as a future requirement \cite{wang2025nonlinear}. Duong et al. construct denoising collision operators that preserve lattice symmetries while approximating the equilibrium manifold \cite{duong2026denoising}.

The present framework differs in the role assigned to the fluid invariants. Encoding-level charge preservation and Gauss-law restoration keep a state inside a physical sector. Here a change in mass and momentum is used as a local relative syndrome at the collision boundary. The D3Q19 velocity set then gives an explicit decoder for a selected charge-changing fault family. The same conservation map exposes a fifteen-dimensional unresolved sector. Exact kinetic streaming resolves that sector by its coupling order to the hydrodynamic variables. This kinetic classification is specific to the streaming structure of LBM and is absent from a pure symmetry check.

\section{Local conservation as a syndrome map}

Consider an LBM stencil with $Q$ populations at a site
\begin{equation}
\vect f=(f_0,\ldots,f_{Q-1})^T.
\end{equation}
The discrete velocities are $\vect c_i$ in $d$ spatial dimensions. For an isothermal fluid the locally conserved collision moments are mass and the $d$ momentum components
\begin{equation}
\rho=\sum_i f_i,
\qquad
j_\alpha=\sum_i c_{i\alpha}f_i,
\qquad
\alpha=1,\ldots,d.
\label{charges}
\end{equation}
Collect the corresponding row vectors in
\begin{equation}
H=
\begin{pmatrix}
1&1&\cdots&1\\
c_{01}&c_{11}&\cdots&c_{Q-1,1}\\
\vdots&\vdots&&\vdots\\
c_{0d}&c_{1d}&\cdots&c_{Q-1,d}
\end{pmatrix}
\in\mathbb{R}^{(d+1)\times Q}.
\label{Hmatrix}
\end{equation}
The isothermal stencils used here then give
\begin{equation}
H_{\mathrm{D2Q9}}\in\mathbb{R}^{3\times 9},
\qquad
H_{\mathrm{D3Q19}}\in\mathbb{R}^{4\times 19}.
\label{Hshapes}
\end{equation}
The local charge vector is
\begin{equation}
\vect q=H\vect f.
\end{equation}
For an ideal unforced collision $\Omega$,
\begin{equation}
H\Omega(\vect f)=H\vect f.
\label{collisionconservation}
\end{equation}
The matrix $H$ holds the exact local collision invariants of the chosen kinetic model. For the isothermal D2Q9 and D3Q19 operators used here those invariants are mass and momentum. An energy-conserving thermal collision would append the conserved energy as one further row. The residual $H\delta f$ remains the syndrome in every such case. The construction depends on that invariant map rather than on the relaxation parameterization. Standard unforced isothermal BGK, two-relaxation-time (TRT), and MRT collisions therefore share the same mass-momentum syndrome, while they differ in how the nonconserved kinetic sector is relaxed.

Let an ideal collision produce $\vect f^\star=\Omega(\vect f)$. A post-collision fault then yields the observed state
\begin{equation}
\vect f_{\mathrm{obs}}^{\star}
=
\Omega(\vect f)+\delta\vect f.
\end{equation}
Charge evaluation gives
\begin{equation}
H\vect f_{\mathrm{obs}}^{\star}
=
H\Omega(\vect f)+H\delta\vect f
=
H\vect f+H\delta\vect f.
\end{equation}
The measured residual is therefore
\begin{equation}
\vect s
=
H\vect f_{\mathrm{obs}}^{\star}-H\vect f
=
H\delta\vect f.
\label{syndromeclassical}
\end{equation}
Equation \eqref{syndromeclassical} is the central classical syndrome relation. It requires the syndrome check to sit between collision and streaming. Local density and momentum change legitimately during streaming. A check after streaming would therefore mix physical transport with fault information.

The decomposition induced by $H$ is also useful. For full row rank $H$, define
\begin{equation}
P_{\parallel}=H^T(HH^T)^{-1}H,
\qquad
P_0=I-P_{\parallel}.
\end{equation}
These projectors satisfy
\begin{equation}
HP_0=0,
\qquad
HP_{\parallel}\delta\vect f=H\delta\vect f.
\end{equation}
Every perturbation can be written as
\begin{equation}
\delta\vect f
=
P_{\parallel}\delta\vect f
+
P_0\delta\vect f.
\label{errordecomposition}
\end{equation}
The first term is the part visible to conservation. The second lies in the kernel (nullspace) of $H$, denoted $\ker H$, and preserves mass and momentum. The two parts require different protection mechanisms.

\subsection{D3Q19 single-population shifts}

For D3Q19 the nineteen velocity vectors consist of the rest direction, six axial directions, and twelve face diagonals. A fault that shifts one population has the form
\begin{equation}
\delta\vect f=\delta\vect e_i,
\end{equation}
where $\vect e_i$ is a population basis vector. Substitution into Eq. \eqref{syndromeclassical} gives
\begin{equation}
\vect s_i(\delta)
=
H(\delta\vect e_i)
=
\delta H\vect e_i.
\end{equation}
The product $H\vect e_i$ selects column $i$ of $H$, so
\begin{equation}
H\vect e_i
=
\begin{pmatrix}
1\\
c_{ix}\\
c_{iy}\\
c_{iz}
\end{pmatrix}
\end{equation}
and
\begin{equation}
\vect s_i(\delta)=
\delta
\begin{pmatrix}
1\\
c_{ix}\\
c_{iy}\\
c_{iz}
\end{pmatrix}.
\label{d3q19signature}
\end{equation}
For $\delta\ne0$,
\begin{equation}
\delta=s_0,
\qquad
\vect c_i=\frac{1}{s_0}(s_x,s_y,s_z)^T.
\label{decoder}
\end{equation}
The D3Q19 velocities are distinct. Equation \eqref{decoder} therefore identifies the population index and the additive shift inside the single-population fault family.

\begin{figure}[t]
\centering
\includegraphics[width=0.72\linewidth]{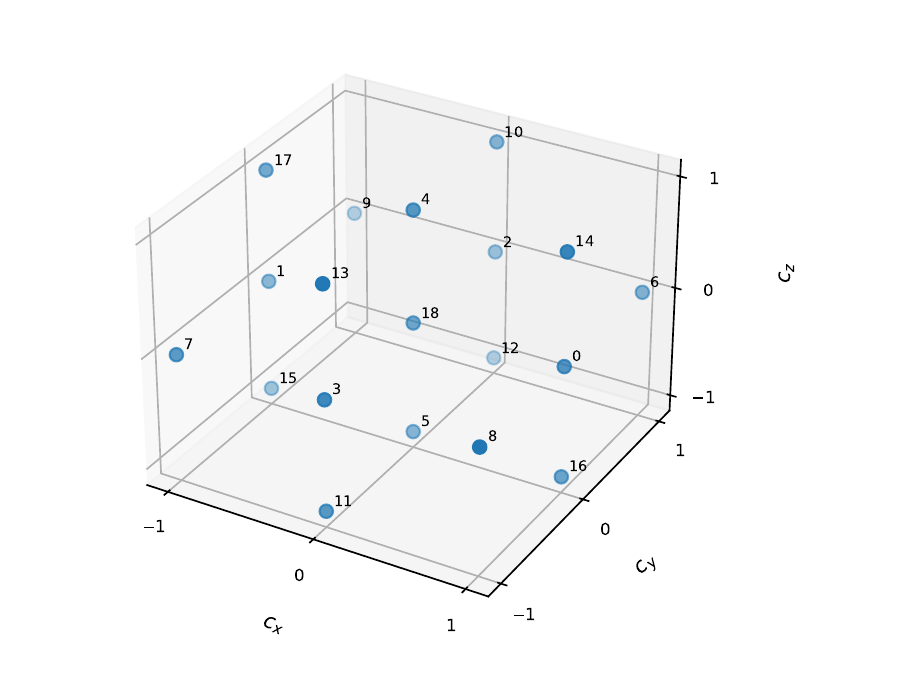}
\caption{D3Q19 velocity directions that label the single-population conservation syndromes. A fault with amplitude $\delta$ produces the four-component signature $\delta(1,\vect c_i)^T$.}
\label{figsyndrome}
\end{figure}

Figure \ref{figsyndrome} shows the nineteen velocity signatures. The syndrome has four components while the direction information occupies the three momentum ratios. The mass component supplies the shift amplitude.

The recovery is direct
\begin{equation}
\vect f_{\mathrm{rec}}=\vect f_{\mathrm{fault}}-s_0\vect e_i.
\label{classicalrecovery}
\end{equation}
This step depends on the assumed error family. A decoder that assumes one additive population shift is aliased by structured multi-fault patterns. The multi-fault analysis in Sec. \ref{limitations} quantifies that boundary.

\section{Quantum reference charges and exact recovery}

Here $\ket{\vect f}$ denotes a computational-basis state whose registers encode the population values. The operator $\hat F_i$ returns the encoded value of population $i$, while $\hat C_a$ collects those values through the corresponding row of $H$. The projector $P_{\vect q}$ selects states with conserved charge $\vect q$, and $\cH_{\vect q}$ denotes the associated joint eigenspace. The operators $K_\ell$ describe the branches of a general quantum collision channel, with a unitary realization recovered as the single-operator special case.

The classical syndrome map becomes a quantum recovery primitive when the conserved populations are represented by commuting observables. Define
\begin{equation}
\hat C_a=\sum_i H_{ai}\hat F_i.
\label{chargeoperators}
\end{equation}
An ideal collision implementation must preserve the relevant charge sectors. For Kraus operators $K_\ell$ of the collision channel, a sufficient branchwise condition is
\begin{equation}
K_\ell P_{\vect q}=P_{\vect q}K_\ell P_{\vect q}.
\label{sectorpreservation}
\end{equation}
The sector condition keeps each collision branch inside one charge eigenspace. Conservation of only the mean mass and momentum would still permit amplitude to occupy other charge values.

A fixed charge sector gives the simplest code picture, yet local fluid charges vary across space and time. We therefore attach a coherent reference register. For computational-basis population states define the isometry
\begin{equation}
V_R\ket{\vect f}_D
=
\ket{\vect f}_D\ket{H\vect f}_R.
\label{referenceisometry}
\end{equation}
The data register $D$ carries the fluid populations. The reference register $R$ carries the local collision invariants. Before collision it can be computed reversibly from the data. Since ideal collision preserves $H\vect f$, the same reference remains valid at the post-collision checkpoint.

An additive population shift on the data register acts as
\begin{equation}
E_{i,\delta}\ket{\vect f}_D
=
\ket{\vect f+\delta\vect e_i}_D.
\end{equation}
On an encoded state the reference register is left unchanged,
\begin{equation}
E_{i,\delta}V_R\ket{\vect f}_D
=
\ket{\vect f+\delta\vect e_i}_D\ket{H\vect f}_R.
\end{equation}
The observed charge of the corrupted data is
\begin{equation}
H(\vect f+\delta\vect e_i)
=
H\vect f+\delta H\vect e_i.
\end{equation}
The relative charge against the stored reference is therefore
\begin{equation}
\Delta\vect q_{i,\delta}
=
H(\vect f+\delta\vect e_i)-H\vect f
=
\delta H\vect e_i
=
\delta\vect h_i,
\label{chargedisplacement}
\end{equation}
where $\vect h_i=H\vect e_i$. Distinct displacements $\Delta\vect q_\mu\ne\Delta\vect q_\nu$ place the two error images in orthogonal relative-charge sectors. With $P=V_RV_R^\dagger$,
\begin{equation}
P E_\mu^\dagger E_\nu P=0
\qquad
\Delta\vect q_\mu\ne\Delta\vect q_\nu.
\label{crosszero}
\end{equation}
For an isometric register shift within its allowed range,
\begin{equation}
P E_\mu^\dagger E_\mu P=P.
\label{diagidentity}
\end{equation}
Hence a selected family with distinct displacements satisfies
\begin{equation}
P E_\mu^\dagger E_\nu P
=\alpha_{\mu\nu}P,
\label{KL}
\end{equation}
with diagonal $\alpha$. Equation \eqref{KL} is the Knill-Laflamme condition \cite{knill1997qec}. The reference register removes the restriction to one fixed physical value of mass and momentum. It preserves the relative syndrome while the logical fluid state spans many charge values.

\subsection{Single population-register bit flips}

A hardware-relevant corollary follows for binary number registers. Write one encoded population as
\begin{equation}
\hat F_i=\sum_{b=0}^{B-1}w_b\hat n_{ib},
\end{equation}
where $\hat n_{ib}$ has eigenvalues zero and one. The weights $w_b$ include the fixed-point scale and the signed weight when a signed binary convention is used. A Pauli $X_{ib}$ flips the stored bit $n_{ib}\mapsto 1-n_{ib}$ and changes the represented population by
\begin{equation}
\delta_{ib}
=(1-2n_{ib})w_b.
\label{bitdelta}
\end{equation}
A stored bit value $n_{ib}=0$ produces $\delta_{ib}=+w_b$. A stored bit value $n_{ib}=1$ produces $\delta_{ib}=-w_b$. The relative charge shift is $\Delta\vect q=\delta_{ib}\vect h_i$ and therefore occupies the two-element set
\begin{equation}
\mathcal S_{ib}
=
\left\{+w_b\vect h_i,-w_b\vect h_i\right\}.
\label{bitset}
\end{equation}
For D3Q19,
\begin{equation}
\vect h_i=(1,\vect c_i)^T.
\end{equation}
The velocity vectors are distinct. Positional binary weights with pairwise-distinct magnitudes, including two's-complement fixed point, satisfy
\begin{equation}
\mathcal S_{ib}=\mathcal S_{jc}
\quad\Longrightarrow\quad
i=j,
\qquad
b=c.
\label{bitinjective}
\end{equation}
Let $\Pi_{+ib}$ and $\Pi_{-ib}$ project onto the two relative-charge sectors in Eq. \eqref{bitset}. Define the degenerate syndrome projector
\begin{equation}
\Pi_{ib}=\Pi_{+ib}+\Pi_{-ib}.
\label{coarseprojector}
\end{equation}
The projectors $\Pi_{ib}$ are mutually orthogonal for distinct labels. Syndrome extraction therefore resolves the coarse label $(i,b)$ while retaining both signs inside the same recovery subspace. The conditioned recovery applies $X_{ib}$ again. The original bit value stays inside the quantum branch rather than entering the classical syndrome record.

For this error family the reference code gives
\begin{equation}
P X_{ib}^\dagger X_{jc}P
=
\delta_{ij}\delta_{bc}P,
\qquad
P X_{ib}P=0.
\label{bitKL}
\end{equation}
Thus the single-bit family is exactly correctable in this idealized number-register model. A D3Q19 encoding with $B$ population bits has
\begin{equation}
N_X=19B
\end{equation}
separate single-bit labels. For illustration, the numerical audit uses a $B=27$ population register, giving $19\times27=513$ single-bit labels and 513 distinct coarse syndrome sets. The choice $B=27$ is a finite register-width example; the injectivity result is formulated for a generic $B$-bit register.

\subsection{Relative syndrome extraction}

The reference charge can be computed with a reversible circuit
\begin{equation}
U_H\ket{\vect f}\ket{0}
=
\ket{\vect f}\ket{H\vect f}.
\end{equation}
The reference is prepared before collision. After collision and a possible fault, the observed charge is computed into a second register. Reversible subtraction gives
\begin{equation}
\ket{\vect q_{\mathrm{ref}}}
\ket{\vect q_{\mathrm{obs}}}
\longrightarrow
\ket{\vect q_{\mathrm{ref}}}
\ket{\vect s},
\qquad
\vect s=\vect q_{\mathrm{obs}}-\vect q_{\mathrm{ref}}.
\label{coherentsyndrome}
\end{equation}
For additive shifts the syndrome directly selects the inverse shift. For binary bit flips a reversible decoder maps the pair of signed values in Eq. \eqref{bitset} to the same coarse label $(i,b)$. The recovery applies $X_{ib}$. The signed residual can then be erased coherently from the recovered data and the stored reference. The coarse label carries the error identity while avoiding a measurement of the original bit value. The abstract correctness follows from Eq. \eqref{bitKL}.

The syndrome layer belongs between collision and streaming. Local density and momentum change during streaming through physical transport. The pre-collision reference is therefore valid for the local collision checkpoint and must be refreshed after the subsequent streaming step.

\begin{figure}[t]
\centering
\includegraphics[width=0.94\linewidth]{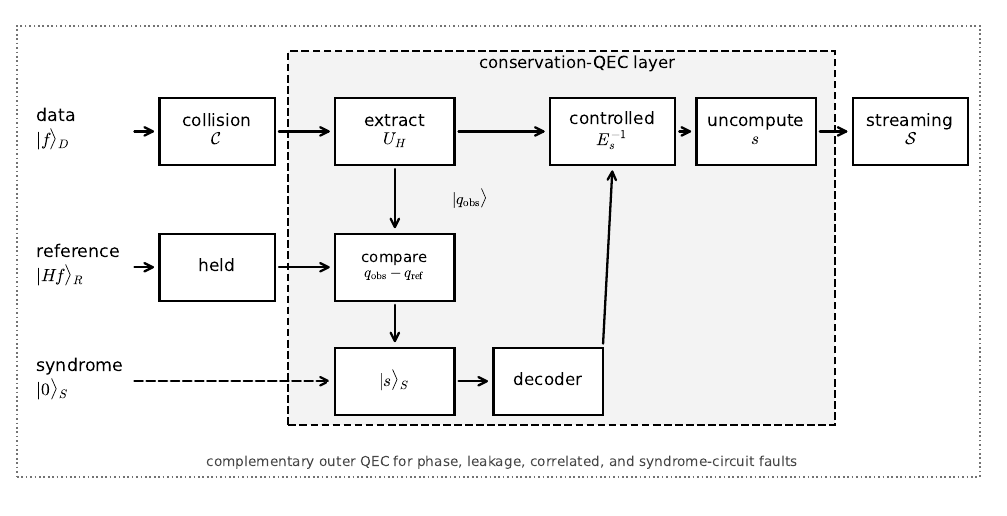}
\caption{One coherent number-register realization of the conservation-QEC layer for LBM. The pre-collision conserved charge is retained coherently in a reference register. After collision, the observed charge is extracted and compared with the reference to form the relative syndrome $s=q_{\mathrm{obs}}-q_{\mathrm{ref}}$. A reversible decoder controls the corresponding recovery. The syndrome is then uncomputed before streaming. Generic outer QEC remains responsible for phase errors, leakage, correlated faults, and faults in syndrome circuitry.}
\label{figframework}
\end{figure}

\subsection{Finite quantum checks}

D2Q9 is the explicit quantum demonstration. Its nine distinct velocities give the local syndrome
\begin{equation}
s_i(\delta)
=
\delta
\begin{pmatrix}
1\\
c_{ix}\\
c_{iy}
\end{pmatrix}.
\end{equation}
The same stencil also contains the in-plane shear mode and the remaining kinetic directions of $\ker H$. Two D2Q9 calculations test the quantum construction directly. The first uses nine two-bit population registers and the interior charge sector
\begin{equation}
\rho=14,
\qquad
j_x=0,
\qquad
j_y=0.
\end{equation}
Ten interior computational-basis states with each population in $\{1,2\}$ span the tested code subspace. The eighteen unit shifts have distinct mass-momentum syndromes. A conservation-syndrome decoder recovers a random logical state to double-precision roundoff. Direct evaluation of every projected product $V^\dagger E_\mu^\dagger E_\nu V$ gives zero off-diagonal residual and zero diagonal error in double precision.

The second calculation tests varying physical charges. Nine two-bit population registers span all $4^9=262144$ D2Q9 data basis states on an $18$-qubit data register. Equation \eqref{referenceisometry} appends the exact mass-momentum reference for each state. The error family contains all eighteen single Pauli $X$ faults across the nine two-bit registers. Their coarse syndrome sets are distinct. Direct Hilbert-space evaluation of every projected product $V_R^\dagger X_\mu^\dagger X_\nu V_R$ gives
\begin{equation}
\max_{\mu\ne\nu}
\left\|V_R^\dagger X_\mu^\dagger X_\nu V_R\right\|_\infty=0,
\end{equation}
\begin{equation}
\max_\mu
\left\|V_R^\dagger X_\mu^\dagger X_\mu V_R-I\right\|_\infty=0,
\end{equation}
and zero overlap with the identity error subspace. Recovery uses the conservation decoder rather than a preassigned inverse. A random complex superposition over all $262144$ logical basis states is recovered to double-precision roundoff for every single bit fault. Equation \eqref{bitKL} covers the linear span of the selected error operators. For a coherent fault operator
\begin{equation}
E=\alpha_0 I+\sum_{i,b}\alpha_{ib}X_{ib},
\end{equation}
the orthogonal syndrome sectors separate the error branches and the conditioned inverse maps every branch to the same logical state. That span statement follows from the Knill-Laflamme relations. The twelve-sample numerical figure is a weighted combination of the already computed single-error recovery fidelities, including the identity branch. It is a derived check of those weights rather than twelve independent simulations of simultaneous coherent faults. Measuring the signed residual would distinguish the original bit value and therefore belongs outside the coherent recovery path. Table \ref{tabsmoke} summarizes the quantum and D3Q19 algebraic checks.

\begin{table}[t]
\caption{D3Q19 algebraic enumerations and the D2Q9 quantum recovery demonstration.}
\label{tabsmoke}
\centering
\begin{tabular}{lr}
\toprule
Quantity & Result \\
\midrule
$\mathrm{rank}(H)$ & 4 \\
$\dim\ker H$ & 15 \\
Audited D3Q19 additive shifts & 304 \\
Distinct additive-shift syndromes & 304 \\
D3Q19 single-bit labels at $B=27$ & 513 \\
Distinct bit-signature sets & 513 \\
Minimum invisible equal-unit support & 4 \\
Minimum invisible multiamplitude support & 3 \\
Fixed-charge D2Q9 code dimension & 10 \\
Fixed-charge maximum QEC residual & 0.0e+00 \\
Reference-code D2Q9 logical dimension & 262144 \\
Reference-code D2Q9 data qubits & 18 \\
Reference-code bit-flip labels & 18 \\
Reference-code maximum QEC residual & 0.0e+00 \\
Minimum recovery fidelity & 1.000000 \\
Minimum weighted single-error fidelity & 0.9999999999999998 \\
\bottomrule
\end{tabular}
\end{table}

\section{The conservation nullspace}
\label{nullspace}

Conservation checks resolve only the component visible to $H$. For D3Q19,
\begin{equation}
H\in\mathbb R^{4\times19},
\qquad
\rank H=4,
\qquad
\dim\ker H=15.
\label{nullity}
\end{equation}
The fifteen-dimensional nullspace contains physically important kinetic moments. Those charge-preserving directions remain unresolved by the conservation syndrome. Hydrodynamic visibility classifies them by the order at which they return to density or momentum. The invariant-syndrome recovery map is independent of the relaxation parameterization. The visibility hierarchy developed below is the explicit D3Q19 MRT realization of that kinetic classification. Other collision models require the corresponding kinetic decomposition.

The MRT basis supplies that classification directly. Let
\begin{equation}
\vect m=M\vect f
\end{equation}
be the D3Q19 moment vector. This basis contains density, three momenta, stress moments, energy-flux-like moments, and higher kinetic moments. Collision acts locally in this basis. Streaming acts diagonally in population space. Its Fourier representation in moment space is
\begin{equation}
T(\vect k)
=
M
\operatorname{diag}
\left(
\exp[-i\vect k\cdot\vect c_0],\ldots,
\exp[-i\vect k\cdot\vect c_{18}]
\right)
M^{-1}.
\label{streamfourier}
\end{equation}
Each streaming phase admits the expansion
\begin{equation}
\exp[-i\vect k\cdot\vect c_i]
=
1
-i(\vect k\cdot\vect c_i)
-\frac12(\vect k\cdot\vect c_i)^2
+\frac{i}{6}(\vect k\cdot\vect c_i)^3
+\cdots.
\end{equation}
Let $P_h$ select the hydrodynamic rows $\rho,j_x,j_y,j_z$ and let $P_r$ select one MRT mode. Insertion into $T(\vect k)$ then gives
\begin{equation}
P_hT(\vect k)P_r
=
\sum_{p=0}^{\infty}
\frac{(-i)^p}{p!}
P_hM
\operatorname{diag}
\left[(\vect k\cdot\vect c_i)^p\right]
M^{-1}P_r.
\label{streamseries}
\end{equation}
Write $\vect k=k\hat{\vect k}$ with magnitude $k=|\vect k|$ and direction $\hat{\vect k}$. Then
\begin{equation}
P_hT(k\hat{\vect k})P_r
=
\sum_{p=0}^{\infty}
k^p A_r^{(p)}(\hat{\vect k}).
\label{streamseriesA}
\end{equation}
The hydrodynamic visibility order is
\begin{equation}
p_r
=
\min\bigl\{p\,\big|\, A_r^{(p)}(\hat{\vect k})\textrm{ fails to vanish identically}\bigr\}.
\label{visibilityorder}
\end{equation}
A coefficient may vanish along a special wave-vector direction while remaining a nonzero function of $\hat{\vect k}$. The order $p_r$ is therefore a generic directional statement.

The same order controls a perturbation that enters immediately before MRT collision. The equilibrium moments depend only on the conserved variables. Hence a kinetic perturbation with $H\delta f=0$ gives $\delta m^{eq}=0$. Linearized MRT collision acts as
\begin{equation}
\delta\vect m^\star=(I-S)\delta\vect m.
\label{kineticcollision}
\end{equation}
For one MRT eigenmode $r$,
\begin{equation}
(I-S)P_r
=(1-s_r)P_r
=\lambda_r P_r,
\end{equation}
and therefore
\begin{equation}
P_hT(\vect k)(I-S)P_r
=\lambda_r P_hT(\vect k)P_r.
\label{collisionvisibility}
\end{equation}
A finite nonzero $\lambda_r$ changes the coefficient while preserving the leading power of $k$. A vanishing $\lambda_r$ removes that mode at collision. A fault inserted after collision is governed directly by Eq. \eqref{streamseries}.

The visibility order measures how quickly a conservation-preserving population perturbation enters density or momentum in the resolved limit. A lower order produces a stronger long-wave hydrodynamic coupling. A higher order gains stronger scale suppression at small $|\vect k|$. Pure phase faults lie outside this population-perturbation classification and belong to the outer quantum code.

\section{Exact D3Q19 visibility hierarchy}

The expansion in Eq. \eqref{streamseries} was evaluated symbolically using the exact integer D3Q19 MRT matrix of this operator. The result separates the kinetic moments into three levels. The energy-like kinetic moment $e$ and the stress sector appear at first order. The energy-flux-like and third-order moments appear at second order. The fourth-order moments appear at third order. Table \ref{tabvisibility} lists every mode.

\begin{table}[t]
\caption{Hydrodynamic visibility of the D3Q19 MRT moments. The order is the first power of $|\mathbf{k}|$ that transfers a moment perturbation into density or momentum during streaming.}
\label{tabvisibility}
\centering
\begin{tabular}{cl}
\toprule
Order & MRT moments \\
\midrule
$O(k^0)$ & $\rho$, $j_x$, $j_y$, $j_z$ \\
$O(k^1)$ & $e$, $p_{xx}$, $p_{ww}$, $p_{xy}$, $p_{yz}$, $p_{xz}$ \\
$O(k^2)$ & $q_x$, $q_y$, $q_z$, $m_x$, $m_y$, $m_z$ \\
$O(k^3)$ & $\epsilon$, $\pi_{xx}$, $\pi_{ww}$ \\
\bottomrule
\end{tabular}
\end{table}

Representative expressions make the hierarchy explicit. For the energy-like kinetic moment $e$,
\begin{equation}
\delta j_x
=-\frac{i k_x}{57}\delta e+O(k^2),
\qquad
\delta j_y
=-\frac{i k_y}{57}\delta e+O(k^2),
\qquad
\delta j_z
=-\frac{i k_z}{57}\delta e+O(k^2).
\label{eleading}
\end{equation}
For the $q_x$ mode,
\begin{align}
\delta j_x
&=-\frac{k_y^2+k_z^2}{20}\delta q_x+O(k^3),\\
\delta j_y
&=-\frac{k_xk_y}{10}\delta q_x+O(k^3),\\
\delta j_z
&=-\frac{k_xk_z}{10}\delta q_x+O(k^3).
\label{qxleading}
\end{align}
Along a wave vector with $k_y=k_z=0$, those second-order couplings vanish. The mode remains an $O(k^2)$ class because $A_{q_x}^{(2)}(\hat{\vect k})$ fails to vanish identically.

For the higher-order energy-like kinetic moment $\epsilon$,
\begin{align}
\delta j_x
&=\frac{i k_x(k_y^2+k_z^2)}{126}\delta\epsilon+O(k^4),\\
\delta j_y
&=\frac{i k_y(k_x^2+k_z^2)}{126}\delta\epsilon+O(k^4),\\
\delta j_z
&=\frac{i k_z(k_x^2+k_y^2)}{126}\delta\epsilon+O(k^4).
\label{epsleading}
\end{align}
These expressions come directly from the D3Q19 streaming matrix. They are independent of the later digital fixed-point implementation.

\begin{figure}[t]
\centering
\includegraphics[width=0.78\linewidth]{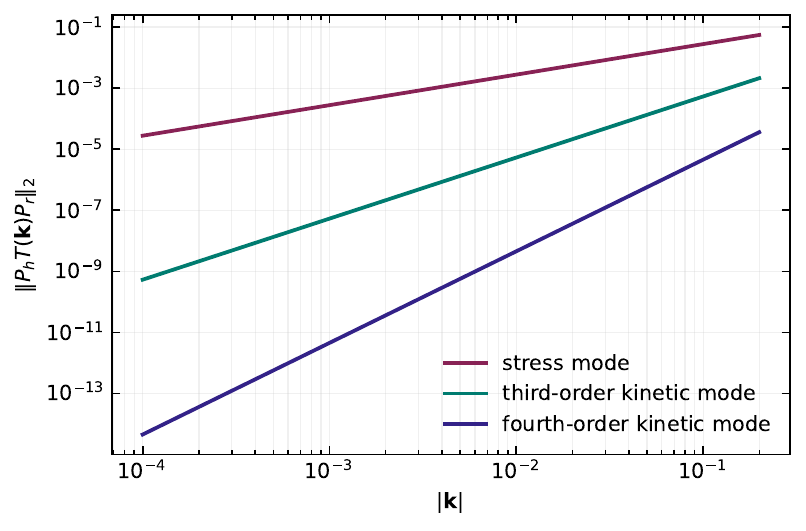}
\caption{Hydrodynamic visibility of representative D3Q19 MRT moments. The stress representative scales as $k$, the third-order kinetic representative scales as $k^2$, and the fourth-order representative scales as $k^3$ in the resolved limit.}
\label{figvisibility}
\end{figure}

A numerical low-wave-number experiment provides a second check. We evaluate
\begin{equation}
\left\|P_hT(k\hat{\vect k})P_r\right\|_2
\end{equation}
for thirty logarithmically spaced values of $k$ and fit the leading slope. The fitted slopes are $0.999993$ for $e$, $1.999995$ for $q_x$, and $2.999994$ for $\epsilon$. Every D3Q19 kinetic mode agrees with its exact integer order to within $10^{-5}$ in the fitted exponent. Figure \ref{figvisibility} shows representative curves.

The visibility hierarchy classifies the conservation nullspace. Errors with $H\delta f\ne0$ enter the conservation decoder. Errors with $H\delta f=0$ remain in the kinetic nullspace. The visibility hierarchy distinguishes those conservation-preserving perturbations according to the order at which they return to the hydrodynamic variables. This information provides a physics-based characterization of the conservation nullspace.

\section{Numerical methods}

\subsection{D3Q19 MRT model and decaying-flow setup}

The numerical experiments use the D3Q19 MRT basis and collision operator of the validated reference solver \cite{khan2026d3q19benchmark}. The velocity ordering and moment matrix are taken from that implementation. The relaxation spectrum is
\begin{equation}
S=
\operatorname{diag}
\left(
0,1.19,1.4,0,1.2,0,1.2,0,1.2,
\frac{1}{\tau},1.4,\frac{1}{\tau},1.4,
\frac{1}{\tau},\frac{1}{\tau},\frac{1}{\tau},1.98,1.98,1.98
\right).
\label{rates}
\end{equation}
Collision is
\begin{equation}
\vect m^\star
=
\vect m-S(\vect m-\vect m^{eq}).
\label{mrtcollision}
\end{equation}
An independent floating-point implementation of this collision operator is used for the deterministic experiments. Direct comparison with the validated solver on a three-dimensional nonequilibrium field gives a maximum population difference of $9.02\times10^{-17}$ and a relative $L_2$ difference of $1.86\times10^{-16}$.

The classical validation uses low-Mach, approximately incompressible, unforced decaying homogeneous isotropic flow on a periodic cube. The classical validation uses low-Mach, approximately incompressible, unforced decaying homogeneous isotropic flow on a periodic cube. Its amplitude is shaped by
\begin{equation}
A(k)=
\left(\frac{k}{k_0}\right)^2
\exp\left[-\left(\frac{k}{k_0}\right)^2\right].
\end{equation}
The inverse transform is normalized to a prescribed total root-mean-square (rms) velocity. Equilibrium D3Q19 populations initialize the flow.

The viscosity is
\begin{equation}
\nu=\frac{\tau-1/2}{3}.
\end{equation}
The initial dissipation is evaluated from the spectral strain tensor
\begin{equation}
\varepsilon
=2\nu\left\langle S_{\alpha\beta}S_{\alpha\beta}\right\rangle.
\end{equation}
With the one-component rms velocity $u'=u_{\mathrm{rms}}/\sqrt{3}$, the Taylor scale and Reynolds number are \cite{pope2000turbulent}
\begin{equation}
\lambda
=
\sqrt{\frac{15\nu u'^2}{\varepsilon}},
\qquad
Re_\lambda
=
\frac{u'\lambda}{\nu}.
\end{equation}
The Kolmogorov length and grid-resolution indicator are
\begin{equation}
\eta
=\left(\frac{\nu^3}{\varepsilon}\right)^{1/4},
\qquad
k_{\max}\eta=\pi\eta
\end{equation}
for unit lattice spacing with the axial Nyquist value $k_{\max}=\pi$.

The main simulation uses $28^3$ sites, $\tau=0.51$, $k_0=1.5$, initial $u_{\mathrm{rms}}=0.05$, and 80 timesteps. Across four initial conditions the starting values are
\begin{equation}
Re_\lambda=43.4\pm0.8,
\qquad
k_{\max}\eta=1.215\pm0.011,
\end{equation}
with total rms Mach number $0.0866$. The spectral divergence rms is about $2.0\times10^{-17}$.

At each step the reference state receives ideal MRT collision and periodic streaming. The two faulty branches receive the same collision. Each site receives a selected additive population fault with probability $0.01$. A selected site receives one randomly chosen D3Q19 population shift with amplitude $\pm2\times10^{-4}$. Faults are inserted after collision and before streaming. The corrected branch stores the expected local charge from the pre-collision state, evaluates the relative syndrome, applies the inverse population shift, and then streams. The uncorrected branch receives the matched fault sequence.

The monitored quantities are the relative velocity error
\begin{equation}
\varepsilon_u
=
\frac{\|\vect u-\vect u_{\mathrm{ref}}\|_2}
{\|\vect u_{\mathrm{ref}}\|_2}
\end{equation}
and the relative shell-energy-spectrum error
\begin{equation}
\varepsilon_E
=
\frac{\|E(k)-E_{\mathrm{ref}}(k)\|_2}
{\|E_{\mathrm{ref}}(k)\|_2}.
\end{equation}
Four independent initial conditions are used.

\section{Results}

\subsection{Syndrome injectivity and failure envelope}

The D3Q19 charge matrix has rank four and nullity fifteen. We enumerate 304 single additive shifts built from all nineteen populations and integer amplitudes $\delta\in\{-8,\ldots,-1,1,\ldots,8\}$. Every shift produces a distinct syndrome. This agrees with Eq. \eqref{decoder}.

For a generic $B$-bit number register, the single Pauli $X$ family contains $19B$ labels. The $B=27$ enumeration therefore contains 513 labels. Equation \eqref{bitset} gives 513 distinct coarse syndrome sets. This is an algebraic consequence of the distinct D3Q19 velocities and distinct binary weight magnitudes.

The equal-unit two-population enumeration contains 684 signed patterns over distinct populations. Every pattern changes at least one conserved quantity. Their syndromes collapse to 254 distinct values. Hence the conservation residual detects this restricted pair family while unique pair decoding needs extra information.

The smallest equal-unit pattern in $\ker H$ has support four. One example is
\begin{equation}
-\vect e_{+x}-\vect e_{-x}
+\vect e_{+y}+\vect e_{-y}.
\label{invisiblefour}
\end{equation}
Allowing unequal amplitudes reduces the minimum support. A support-three example is
\begin{equation}
-\vect e_{+x}-\vect e_{-x}+2\vect e_0.
\label{invisiblethree}
\end{equation}
Both expressions preserve mass and momentum exactly and therefore enter the hydrodynamic-visibility layer.

Multi-fault aliases also limit a decoder designed for one additive population shift. The equal-unit pair analysis identifies 42 patterns whose syndrome matches a larger single shift. For example
\begin{equation}
\vect e_{+x}+\vect e_{-x}
\end{equation}
shares the syndrome of $2\vect e_0$. A multiscale additive-pair alphabet with amplitudes $\pm1$, $\pm2$, $\pm4$, and $\pm8$ contains 378 pair patterns that alias one allowed single shift. These counts bound that additive alphabet. Pauli-$X$ bit-fault pairs form a different family and require a separate alias analysis.

\subsection{D3Q19 decaying-flow recovery}

The main simulation injects $17531 \pm 183$ selected single-shift faults per realization. Every injected fault from that alphabet is recovered in the classical simulation. The largest post-recovery local charge residual is $(6.80\pm0.28)\times10^{-16}$ and is set by floating-point roundoff.

After 80 steps, the uncorrected branch reaches
$\varepsilon_u=(1.674\pm0.013)\times10^{-3}$,
whereas the recovered branch reaches
$\varepsilon_u=(4.467\pm0.035)\times10^{-15}$,
which is at the level of classical floating-point roundoff.

\begin{figure}[t]
\centering
\includegraphics[width=0.78\linewidth]{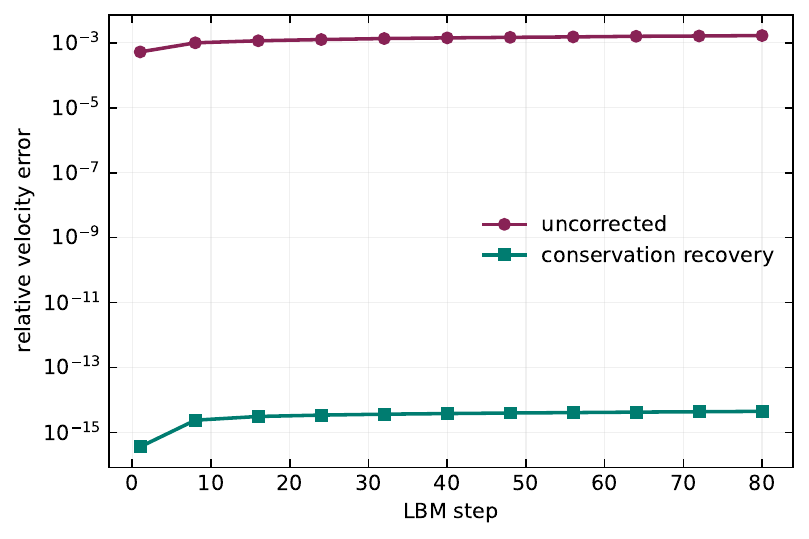}
\caption{Relative velocity error in the D3Q19 MRT decaying-flow simulation. Curves show the four-seed mean for matched injected faults. Conservation recovery returns the selected fault family to the reference trajectory within floating-point roundoff.}
\label{figdhit}
\end{figure}

The uncorrected spectrum error is
$\varepsilon_E=(7.19\pm2.37)\times10^{-5}$,
while the recovered spectrum error is
$\varepsilon_E=(3.20\pm1.11)\times10^{-16}$.
Table \ref{tabdhit} gives the complete summary. Figure \ref{figdhit} shows the velocity error through time.

\begin{table}[t]
\caption{D3Q19 MRT decaying homogeneous isotropic turbulence simulation with injected faults. Values are mean $\pm$ one standard deviation over 4 seeds.}
\label{tabdhit}
\centering
\begin{tabular}{lr}
\toprule
Quantity & Result \\
\midrule
Injected faults per realization & $17531 \pm 183$ \\
Corrected fraction & $1.000000 \pm 0.000000$ \\
Uncorrected velocity error & $(1.674 \pm 0.013)\times10^{-3}$ \\
Recovered velocity error & $(4.467 \pm 0.035)\times10^{-15}$ \\
Uncorrected spectrum error & $(7.188 \pm 2.373)\times10^{-5}$ \\
Recovered spectrum error & $(3.196 \pm 1.109)\times10^{-16}$ \\
\bottomrule
\end{tabular}
\end{table}

The exact recovery seen here follows from the algebraic decoder. The flow experiment serves a different purpose from the injectivity proof. It verifies repeated syndrome extraction and recovery while the local populations evolve through a three-dimensional broadband field over many collision-streaming steps.

\begin{figure}[t]
\centering
\includegraphics[width=0.78\linewidth]{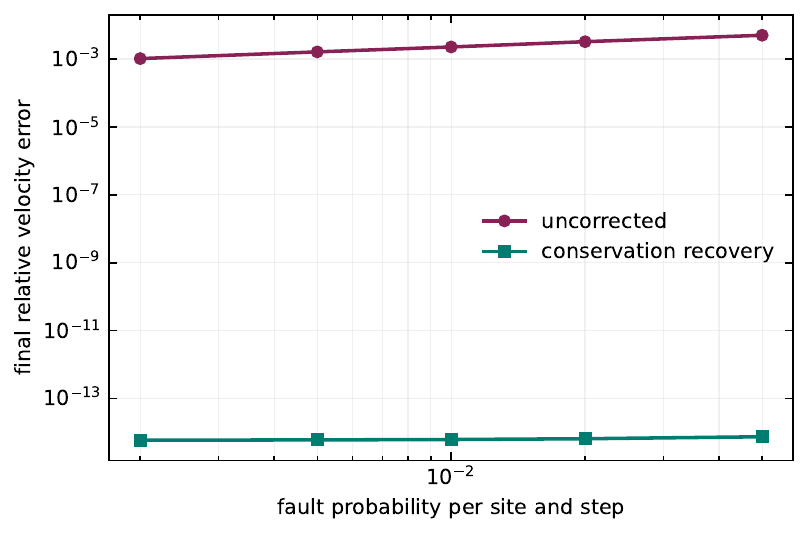}
\caption{Final velocity error as a function of the per-site fault probability. The recovery curve stays at numerical roundoff for the selected single-population fault alphabet while the uncorrected branch grows with fault rate.}
\label{figsweep}
\end{figure}

To examine sensitivity to the frequency of injected faults, a separate $16^3$ simulation uses 50 steps and varies the per-site fault probability from $0.002$ to $0.05$. The highest setting produces about $1.03\times10^4$ selected single-shift faults per realization. Every injected fault from that alphabet remains recovered over the full range of fault probabilities. The uncorrected final velocity error rises from about $1.03\times10^{-3}$ to $5.05\times10^{-3}$. The recovered error remains between $5.86\times10^{-15}$ and $7.41\times10^{-15}$. Figure \ref{figsweep} shows the separation.

\section{Implications for quantum LBM design}

The results define an inner recovery construction for quantum LBM whose first component uses exact collision invariants. A charge-changing error produces a local relative syndrome, and for the selected D3Q19 single-population shift family this map is injective, allowing the error identity to be resolved from the syndrome. Binary number registers provide a second exact family, since each single bit flip has a distinct coarse label determined by its population velocity direction and binary weight.

The second component characterizes the conservation nullspace. Population-value errors in $\ker H$ preserve mass and momentum and therefore remain invisible to the conservation-syndrome layer, yet their subsequent hydrodynamic influence is not uniform. Stress-like modes enter the conserved variables at $O(k)$, third-order kinetic modes at $O(k^2)$, and fourth-order modes at $O(k^3)$. Collision introduces an additional mode-dependent factor through $\lambda_r=1-s_r$, so persistence after collision is governed by $|\lambda_r|$, while hydrodynamic return is governed by the visibility order. Equation \eqref{collisionvisibility} contains both effects and therefore distinguishes conservation-preserving perturbations according to both their collision persistence and the order at which they return to the hydrodynamic variables.

The third component is a generic outer quantum code for mechanisms that fall outside the selected charge-shift families, including phase errors, leakage, correlated faults, and faults in syndrome arithmetic. A stabilizer code or another fault-tolerant architecture can provide protection against these mechanisms, while conservation recovery remains an LBM-specific inner layer rather than a replacement for general fault tolerance.

Beyond exact collision invariants, particular LBM formulations may also provide lattice isotropy, Galilean or covariance properties, entropy constraints, and stability information \cite{lallemand2000theory,kruger2017lbm,karlin1999entropic}. Those statements differ in kind from a local collision equality. Rotational symmetry belongs to the velocity set \cite{lallemand2000theory}, while a sheared fluid state may occupy unequal opposite populations. Finite-velocity Galilean residuals already sit in the kinetic truncation \cite{lallemand2000theory,dhumieres2002mrt}, so a frame mismatch can come from the model itself. An entropic $H$-function inequality belongs to entropic collisions \cite{karlin1999entropic,ansumali2003minimal}, whereas the present operator is MRT. Covariance compares a transformed evolution with an evolved transform, which is a map-level test rather than a local charge residual \cite{kruger2017lbm}. Those properties can support verification, admissibility checks, or implementation tests. The active syndrome stays attached to exact collision invariants.

For number-register encodings, the conserved charges can be accumulated with reversible arithmetic, making the reference-syndrome construction directly realizable in principle. Amplitude-encoded QLBM requires a different realization of the conserved observables, but the same algebraic syndrome principle remains applicable whenever the charge sectors and their relative changes can be extracted coherently. The underlying LBM charge map is therefore independent of the quantum encoding, even though its circuit realization depends on how the population data are represented.

\section{Limits and falsification tests}
\label{limitations}

The active recovery result applies to a stated error family. The additive decoder assumes a single population shift at a local collision checkpoint, while multi-fault patterns can share the same syndrome, as shown by the alias analysis. For binary number registers, the reference-charge construction gives exact recovery for the selected single-$X$ family. Correlated bit faults fall outside that family and require an expanded decoder or protection by an outer code.

Pure phase faults preserve the encoded population values and therefore leave the four conservation charges unchanged, making them invisible to the conservation syndrome. The hydrodynamic-visibility analysis likewise concerns population-value perturbations rather than pure quantum phase errors. Protection against phase faults therefore remains a task for the outer code.

The syndrome reference must itself be reliable, since a single corrupted reference bit can send an unfaulted data register into a nonzero relative syndrome. The decoder may then apply a spurious inverse even though the residual against the corrupted reference subsequently vanishes. A check based only on $Hf_{\mathrm{obs}}-Hf_{\mathrm{ref}}$ therefore cannot detect corruption of that same reference. Verified ancillas, repeated syndrome extraction, or an outer code can protect this circuitry. The present construction establishes the LBM-specific inner recovery conditions rather than a complete hardware fault-tolerant architecture.

Finite-precision arithmetic requires an exactly defined collision checkpoint. A finite-width inverse moment transform can leave a deterministic charge residual even when every computational step is ideal. The inner syndrome therefore belongs at a representation in which the encoded charges are preserved exactly by the ideal arithmetic, for example the conserved MRT moment registers. When a deterministic reconstruction residual can be computed from the ideal arithmetic independently of the observed fault, it may be included in the expected reference. A numerical tolerance would replace the exact orthogonal-syndrome argument by an approximate test. Binary bit flips are unitary permutations on a fixed-width register and avoid additive-shift boundary problems. A direct additive translation requires guard space, modular arithmetic, or an overflow flag. The classical floating-point simulation evaluates conservation to roundoff. A reversible integer implementation can maintain the reference in the chosen exact arithmetic.

Forcing changes the expected momentum during collision. A known source can be incorporated through
\begin{equation}
  \vect s
  =
  H\vect f^{\star}_{\mathrm{obs}}
  -H\vect f_{\mathrm{pre}}
  -\Delta\vect q_{\mathrm{phys}},
  \label{forcedsyndrome}
  \end{equation}
where $\Delta\vect q_{\mathrm{phys}}$ is the prescribed physical source increment. Boundary fluxes can enter the same accounting when their contribution is known. The numerical experiment uses periodic unforced decay so that the local collision reference remains exact. Hydrodynamic visibility should likewise be interpreted as a resolved-scale statement. Near the lattice cutoff, the separation associated with successive powers of $k$ becomes weaker. Equation \eqref{collisionvisibility} supplies the collision contribution, while the exact D3Q19 streaming expansion supplies the scale-dependent contribution.

The strongest falsification tests follow directly from these limitations. The reference-code Knill-Laflamme products must retain the required diagonal structure for the claimed error alphabet, and the D3Q19 bit-label sets must remain injective. The hydrodynamic-visibility exponents must agree with the exact streaming expansion. Any broader hardware noise model must also be tested against its own syndrome degeneracies and alias structure before a corresponding correction claim is made.

\section{Discussion}

The conservation-syndrome view changes the computational role of LBM invariants by using mass and momentum not only to constrain the collision model but also to classify charge-changing faults. The collision-streaming split provides a particularly clean checkpoint because local collision preserves these charges, whereas subsequent streaming changes them through physical transport.

The reference-charge construction is essential for a fluid solver because a fixed charge sector would restrict the logical state to one local value of mass and momentum. Equation \eqref{referenceisometry} instead allows these quantities to vary across sites and timesteps by storing their expected values coherently and diagnosing only the relative change created at the collision checkpoint. The D2Q9 variable-charge state-vector calculation verifies this construction directly for the selected single-bit-flip family.

The distinction from generic symmetry-based QEC lies in how the physical structure is used. Existing symmetry-based codes show that conserved or constrained structure can define protected sectors \cite{niu2018symmetry,qin2024symmetry,bradshaw2026symmetry,rajput2023gauge,spagnoli2026gauge}. Rajput et al. exploit the Gauss-law constraint as part of an error-correction construction for lattice gauge theories. The LBM contribution is the explicit local syndrome geometry $\delta(1,\vect c_i)^T$, its direct D3Q19 decoder, and the kinetic analysis of the fifteen-dimensional charge-preserving sector. Schalkers and M\"oller preserve mass and momentum inside a unitary encoding \cite{schalkers2024encoding}. Fonio et al. study conserved quantum invariants in lattice-gas automata \cite{fonio2025invariants}. Tiwari et al. treat QLBM hardware noise through mitigation \cite{tiwari2025realizable}. The present construction turns the classical collision invariants into an active recovery condition for a selected quantum error family.

Open-system and digital QLBM constructions address how the LBM dynamics themselves can be implemented as valid quantum evolution. In contrast, conservation-syndrome QEC addresses errors that occur while such a quantum LBM computation is being executed. The present recovery construction therefore complements, rather than replaces, quantum implementations of dissipative or nonlinear LBM dynamics. Its reference-code derivation does not depend on the specific two-bank compute-reset architecture, and number-register encoding is used only as one concrete realization of the underlying charge operators.

The hydrodynamic-visibility analysis provides a further distinction by resolving structure within the fifteen-dimensional D3Q19 conservation nullspace. A conservation check leaves these population directions indistinguishable at the charge level, but treating the entire nullspace as one unresolved sector discards important kinetic information. Exact streaming shows that stress-like, third-order, and fourth-order modes return to the macroscopic fields at different powers of wave number. The resulting two-layer classification therefore separates charge-changing faults from charge-preserving population perturbations and, within the latter class, distinguishes modes according to the order at which they re-enter the hydrodynamic variables.

\section{Conclusion}

Local LBM collision invariants can carry quantum error information. An additive post-collision fault produces the relative syndrome $s=H\delta f$, and for D3Q19 a single population shift gives $\delta(1,\vect c_i)^T$. Distinct velocities make the selected single-shift map injective. A coherent reference-charge register extends this construction across varying local mass and momentum by converting conservation changes into relative quantum syndrome sectors. The selected single population-register bit-flip family generates disjoint coarse syndrome sets, and the associated projected error products satisfy the Knill-Laflamme condition. The D2Q9 variable-charge state-vector verification gives zero QEC residual and recovery to double-precision roundoff. For a $B$-bit D3Q19 number representation, the construction supplies $19B$ distinct single-bit labels.

Conservation leaves a fifteen-dimensional D3Q19 kinetic nullspace, and exact MRT streaming analysis resolves its population-value perturbations into $O(k)$, $O(k^2)$, and $O(k^3)$ hydrodynamic visibility classes. MRT relaxation changes the coefficient of a pure kinetic mode while preserving its visibility order whenever the relaxation eigenvalue is finite and nonzero. The D3Q19 decaying-flow simulation then verifies repeated recovery of the selected additive single-shift alphabet in a moderate-Reynolds-number three-dimensional field. About $1.75\times10^4$ selected faults are injected per realization, and the corrected classical-simulation trajectory returns to floating-point roundoff while the matched uncorrected trajectory accumulates measurable velocity and spectral error.

The resulting principle is compact. Exact local collision invariants of the chosen kinetic model supply an active syndrome for charge-changing faults. Kinetic hydrodynamic visibility, together with the collision eigenvalue, distinguishes charge-preserving population faults in the conservation nullspace. Generic quantum codes can provide protection against the remaining hardware mechanisms. In this way the structure of the fluid algorithm participates directly in quantum error correction.

Several extensions follow from this foundation. Correlated faults, phase errors, leakage, and faults in syndrome circuitry require an expanded recovery construction or combination with an outer quantum code. The invariant map can also be studied for other kinetic models and quantum encodings, including thermal LBM with additional conserved quantities and amplitude-encoded QLBM. A further step is an exact finite-arithmetic circuit realization of the conservation checkpoint. Once these elements are established, an equal-resource comparison with generic fault-tolerant QLBM can determine whether physics-aware protection reduces the overall cost of reliable quantum fluid simulation.

\section*{Data and code availability}

\printbibliography

\end{document}